%% file: main.tex
\documentclass[10pt,twocolumn,letterpaper]{article}

\usepackage{cvpr}      

\input{preamble}

\usepackage{graphicx}
\usepackage{newtxtext}
\usepackage{booktabs}
\usepackage{makecell}

\usepackage{tabularx}

\usepackage{comment}

\usepackage{float}

\definecolor{cvprblue}{rgb}{0.21,0.49,0.74}
\usepackage[pagebackref,breaklinks,colorlinks,allcolors=cvprblue]{hyperref}

\usepackage{xcolor} 

\def\paperID{*****} 
\def\confName{CVPR}
\def\confYear{2026}

\title{
GUIDE: Guided Updates for In-context Decision Evolution in LLM-Driven Spacecraft Operations\\
\vspace{0.5em}
{\large Accepted at CVPR 2026 AI4Space Workshop}
}
\author{Alejandro Carrasco\\
Massachusetts Institute of Technology\\
Cambridge, USA\\
{\tt\small acarra@mit.edu}
\and
Mariko Storey-Matsutani\\
Massachusetts Institute of Technology\\
Cambridge, USA\\
{\tt\small m\_storey@mit.edu}
\and
Victor Rodriguez-Fernandez\\
Universidad Politecnica de Madrid\\
Madrid, Spain\\
{\tt\small victor.rfernandez@upm.es}
\and
Richard Linares\\
Massachusetts Institute of Technology\\
Cambridge, USA\\
{\tt\small linaresr@mit.edu}
}
\begin{document}

\maketitle
\input{sec/0_abstract}    
\input{sec/1_intro}
\input{sec/2_background}
\input{sec/3_methodology}
\input{sec/4_results}
\input{sec/5_conclusions_and_future_work}
\input{sec/6_acknowledgements}
{
    \small
    \bibliographystyle{ieeenat_fullname}
    \bibliography{main}
}

\input{sec/X_suppl}


\end{document}

%% file: preamble.tex
\usepackage{multirow}



%% file: sec/0_abstract.tex
\begin{abstract}
Large language models (LLMs) have been proposed as supervisory agents for spacecraft operations, but existing approaches rely on static prompting and do not improve across repeated executions. We introduce \textsc{GUIDE}, a non-parametric policy improvement framework that enables cross-episode adaptation without weight updates by evolving a structured, state-conditioned playbook of natural-language decision rules. A lightweight acting model performs real-time control, while offline reflection updates the playbook from prior trajectories. Evaluated on an adversarial orbital interception task in the Kerbal Space Program Differential Games environment, GUIDE's evolution consistently outperforms static baselines. Results indicate that context evolution in LLM agents functions as policy search over structured decision rules in real-time closed-loop spacecraft interaction.
\end{abstract}

%% file: sec/1_intro.tex
\section{Introduction}

Spacecraft operations remain costly, high-risk, and slow to iterate. Unlike most terrestrial AI deployments, missions cannot easily collect large amounts of new data, reset the environment, or retrain policies after deployment. Yet many operational scenarios—exploration missions, autonomous Guidance, Navigation, and Control (GNC), on-orbit servicing, and non-cooperative interaction—require supervisory reasoning under uncertainty rather than only low-level control. This creates a gap between existing autonomy algorithms and operator-level decision making, motivating systems capable of reasoning under uncertainty and adaptability rather than relying on fixed policies.

Large language models (LLMs) have emerged as promising candidates for this supervisory layer \cite{rodriguezfernandez2024languagemodelsspacecraftoperators, doi:10.2514/6.2025-1543}. Beyond text generation, they exhibit reasoning, hierarchical agentic behavior, and structured decision making \cite{yao2022react}. Modern LLM systems can also alter behavior at inference time through in-context learning and memory without updating weights, making them attractive where retraining is impractical and deployment constraints dominate.

Deploying an LLM as a supervisory decision agent in a real-time, closed-loop dynamical system introduces two coupled constraints. First, control actions must be generated within strict latency bounds while reasoning over delayed feedback, irreversible actuation, and adversarial interactions. Models capable of extended deliberation incur prohibitive latency for direct integration into the control loop, whereas models that meet real-time requirements cannot perform such reasoning during action selection. Second, interaction with reactive or unknown agents requires adaptation across encounters rather than one-shot decisions. Because such adaptation cannot occur inside the online loop, learning must be separated from action execution. This dual constraint motivates a teacher–student separation, in which a frontier reasoning model performs episodic offline reflection while a lightweight acting model executes real-time control.

Building on the paradigm of large language models as spacecraft operators~\cite{rodriguezfernandez2024languagemodelsspacecraftoperators, CARRASCO20253480, doi:10.2514/6.2025-1543}, we introduce GUIDE (\textbf{G}uided \textbf{U}pdates for \textbf{I}n-context \textbf{D}ecision \textbf{E}volution), a non-parametric policy improvement framework that enables cross-episode adaptation without parameter updates. Instead of optimizing parameters, the system iteratively refines a state-conditioned playbook—a structured set of natural language rules encoding contextual behavior guidance—which functions as a non-parametric policy representation. The acting model remains fixed; policy improvement arises from updating this context rather than model weights, enabling adaptation to adversarial dynamics while preserving real-time execution. We show that structured in-context memory can serve as a learnable policy object improved across episodes without parameter updates.

To study this setting, we consider a real-time multi-agent Capture-the-Satellite scenario using the Kerbal Space Program Differential Games (KSPDG) environment \cite{10115968ross}. The environment involves sequential decision-making under delayed observations and non-cooperative interactions, providing a closed-loop setting in which cross-episode adaptation can be evaluated under realistic dynamical constraints.

%% file: sec/2_background.tex
\section{Related Work}
Spacecraft autonomy requires sequential decision-making under delayed observations and irreversible dynamics, particularly in adversarial and non-cooperative scenarios~\cite{harris_spacecraft_2019, mehlman2024cat}. While classical GNC provides reliable stabilization and trajectory tracking when objectives and models are specified~\cite{CARRASCO20253480}, it does not address strategic objective selection. This motivates a supervisory decision layer above conventional control, for which recent work has explored LLMs as operator-like decision modules for spacecraft tasks~\cite{rodriguezfernandez2024languagemodelsspacecraftoperators, CARRASCO20253480, doi:10.2514/6.2025-1543}.

More broadly, LLM-agent research studies systems combining reasoning, memory, and tool use within action loops~\cite{wang2023surveyllmagents, yao2022react}, but these agents accumulate errors and exhibit prompt sensitivity in long-horizon tasks~\cite{wang2023surveyllmagents}. A key observation is that behavior can adapt at inference time through in-context learning (ICL): instead of weight updates, the model conditions on demonstrations and memory to modify its decisions~\cite{codaforno2023metaincontextlearninglargelanguage, dai2023gptlearnincontextlanguage, li2025brewingknowledgecontextdistillation}. This has been interpreted as inference-time meta-learning~\cite{codaforno2023metaincontextlearninglargelanguage}, and reflection mechanisms incorporate experience without gradient-based retraining~\cite{shinn2023reflexion, madaan2023selfrefine, park2023generativeagents}.

Agentic Context Engineering (ACE) treats structured context as the persistent object of improvement across executions~\cite{zhang2026agenticcontextengineeringevolving}, with related self-evolving systems refining structured experience across runs~\cite{zhou2026wiseflow, hu2026controlledselfevolution}. Our framework is closest in spirit to ACE but differs in objective and role: rather than using evolving context solely to guide behavior, the context itself becomes the learned decision representation that shapes the actions of an online agent. Unlike prior self-evolving agents, which are typically evaluated in episodic software environments, we study context evolution within a continuous closed-loop dynamical control system.

%% file: sec/3_methodology.tex
\section{Method}

\begin{figure*}[t]
    \centering
    \includegraphics[width=\linewidth,height=0.285\textheight,keepaspectratio]{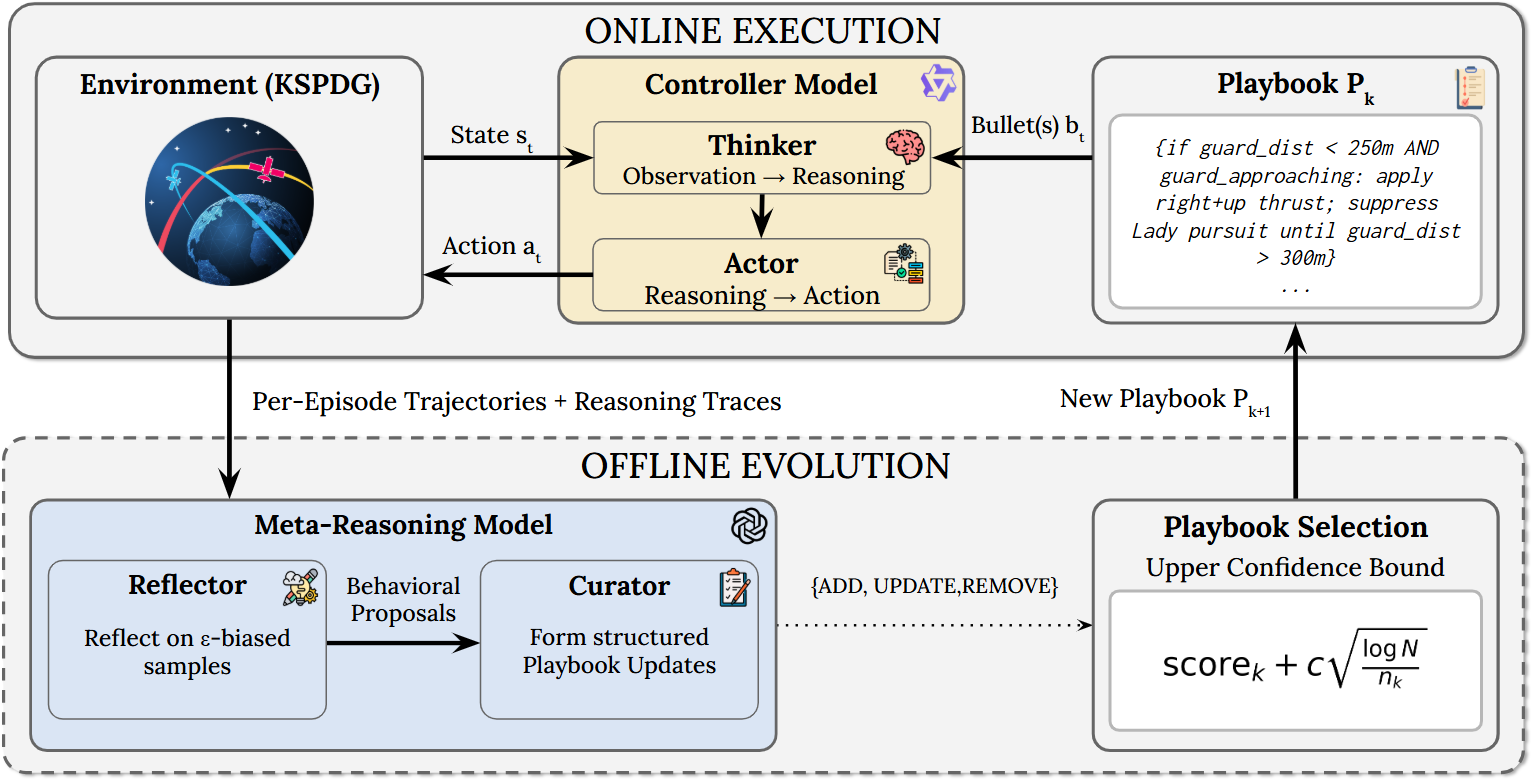}
    \caption{Closed-loop control with cross-episode context adaptation. A fixed acting model executes $\pi(a\mid s,P_k)$ online, while offline reflection updates the playbook via \textsc{add}/\textsc{update}/\textsc{remove} and UCB selects improved versions.}
    \label{fig:diagram}
\end{figure*}

We propose \textsc{GUIDE}, a framework for language-based decision agents in real-time closed-loop tasks (Fig.~\ref{fig:diagram}). The key idea is to learn state-conditioned behavioral guidance expressed in natural language (a \emph{playbook}) and iteratively improve it using episodic reflection over executed trajectories. Instead of updating model parameters, the method performs a search over structured rules, while a resource-constrained language model executes actions online.

\subsection{Problem Setting}

We consider sequential decision problems defined over continuous spacecraft dynamics. At each timestep $t$, the agent observes state $s_t$ and produces a control action $a_t$. State transitions are governed by orbital mechanics, and thrust actions irreversibly influence future trajectories.

Experiments use the KSPDG environment, which simulates adversarial orbital encounters under varying initial conditions and maneuver policies. We evaluate the Lady–Bandit–Guard (LBG) interception scenario, where the agent controls the bandit spacecraft $B$, attempting to minimize its distance to the lady spacecraft $L$ while avoiding capture by the guard spacecraft $G$. At each timestep, the agent observes elapsed time $t$, spacecraft mass ($m_{prop}, m_{total}$), and CBCI-frame position and velocity vectors for all relevant spacecraft, yielding
$$
s_{t, LBG} = \{ t_{LBG}, m_{B,total}, m_{B,prop}, \vec{B}, \dot{\vec{B}}, \vec{L}, \dot{\vec{L}}, \vec{G}, \dot{\vec{G}} \},
$$
where $\vec{B},\dot{\vec{B}}$ denotes the bandit's position and velocity, $\vec{L},\dot{\vec{L}}$ the lady's, and $\vec{G},\dot{\vec{G}}$ the guard's. The control action $a_t$ is a three-axis thrust vector with an associated burn duration~\cite{10115968ross}.

Guard behavior varies across scenarios. In LG4, the Guard follows an aggressive pursuit policy, directly chasing the Bandit. In LG6 and LG7, the Guard adopts a defensive strategy, remaining near the Lady and positioning along likely interception trajectories. LG5 introduces stochastic switching between pursuit and defense. These regimes are not specified in prompts or observations and are not known a priori to the LLM policy generator.

\subsection{Playbook Representation}

Rather than learning a policy $\pi_\theta(a|s)$ through gradient optimization, action selection is conditioned on a language-structured context $\pi(a|s, P)$, where $P$ is a dynamic playbook providing state-conditioned behavioral guidance.

The acting model receives a fixed baseline prompt encoding generic domain instructions (e.g., guard avoidance and prograde alignment for interception). This baseline remains constant across all experiments and is not modified during training.

The playbook $P$ is initially empty and evolves over episodes. It is represented as a collection of bullets or natural language rules, optionally annotated with a state predicate $b_i = (\text{text}_i, \text{condition}_i(s))$, where a bullet becomes active when its condition evaluates to true (e.g., distance thresholds or approach conditions). Active bullets are injected into the prompt at runtime, augmenting the fixed baseline context and modifying the decision distribution.

The learned object in our framework is therefore not the model parameters nor the fixed baseline prompt, but the evolving playbook $P$.

\subsection{Online Decision Agent}

The acting language model executes control decisions at every timestep using the playbook-conditioned policy $\pi(a \mid s, P)$. To ensure real-time compliance while preserving structured reasoning, the model is decomposed into two sequential calls within each control cycle. 

First, a bounded \textbf{reasoning agent} (fixed token budget) produces short-horizon intent conditioned on the current observation and the active playbook bullets. All strategic decision-making occurs within this reasoning step, and the playbook $P$ directly shapes the model’s deliberation through state-conditioned behavioral rules injected into context.

Second, an \textbf{actor agent} maps the state and the generated intent to a structured control output via a function call tool. This stage produces no free-form reasoning and is restricted to the control action format required by the environment.

This separation bounds reasoning latency while allowing the evolving playbook to shape strategic behavior; the actor simply enforces structured execution within the control interface.

\subsection{Offline \texorpdfstring{$\epsilon$}{epsilon}-Biased Reflection Sampling}

After batches of episodes, trajectories are evaluated by the meta-reasoning LLM using task performance metrics and constraint violations. Reflection inputs are not chosen solely from failures. Instead, an $\epsilon$-selection rule is used: with probability $1-\epsilon$, the system reflects on poorly performing trajectories, and with probability $\epsilon$, it reflects on successful ones. Each reflection input consists of a short telemetry window together with the reasoning trace that produced the corresponding decisions.

A frontier language model (\emph{Reflector}) analyzes the selected episode and proposes behavioral corrections. A second pass (\emph{Curator}) converts these proposals into structured operations applied to a rule-based playbook, producing new rules for a new playbook version $P_k$:
\[
\{\text{ADD}, \text{UPDATE}, \text{REMOVE}\},
\]

Multiple playbook versions are maintained simultaneously and evaluated over complete episodes. Selection among playbooks uses an Upper Confidence Bound (UCB1) rule
\[
\text{score}_k + c \sqrt{\frac{\log N}{n_k}},
\]
where $n_k$ is the number of evaluations of version $k$ and $N$ is the total number of evaluations. This favors playbooks that perform well while still allowing newly generated versions to be tried.

The acting model performs real-time control using the current playbook, while the frontier model operates only between episode batches to revise the playbook from accumulated trajectories. The two models, therefore, serve complementary roles: online execution and offline policy improvement.

%% file: sec/4_results.tex
\section{Results \& Discussion}

\begin{table*}[t]
    \centering
    \small
    \setlength{\tabcolsep}{2pt}
    \caption{Composite score comparison across LBG scenarios (lower is better). Mean\,$\pm$\,std. N = 20}
    \label{tab:policy_comparison}
    \begin{tabular}{lcccc}
        \toprule
        Policy & LG4 & LG5 & LG6 & LG7 \\
        \midrule
        LLM (static v0)
            & $4.15{\times}10^{5} \pm 7.72{\times}10^{5}$
            & $5.37{\times}10^{5} \pm 9.30{\times}10^{5}$
            & $2.28{\times}10^{7} \pm 1.84{\times}10^{7}$
            & $1.95{\times}10^{7} \pm 1.28{\times}10^{7}$ \\
        GUIDE (best evolved)
            & $7.22{\times}10^{4} \pm 3.53{\times}10^{4}$
            & $8.49{\times}10^{4} \pm 4.07{\times}10^{4}$
            & $\mathbf{5.07{\times}10^{4}} \pm 2.38{\times}10^{4}$
            & $\mathbf{3.49{\times}10^{4}} \pm 2.33{\times}10^{4}$ \\
        \quad\textit{p-value vs.\ v0}
            & \textit{0.043}
            & \textit{0.143}$^{\dagger}$
            & \textit{${<}0.001$}
            & \textit{${<}0.001$} \\
        LQR $\pi$
            & $1.18{\times}10^{6} \pm 8.12{\times}10^{5}$
            & $9.87{\times}10^{5} \pm 9.02{\times}10^{5}$
            & $4.78{\times}10^{5} \pm 1.82{\times}10^{5}$
            & $1.74{\times}10^{5} \pm 4.90{\times}10^{4}$ \\
        Prograde-Alignment $\pi$
            & $\mathbf{3.73{\times}10^{4}} \pm 6.97{\times}10^{4}$
            & $\mathbf{1.60{\times}10^{4}} \pm 1.44{\times}10^{4}$
            & $2.92{\times}10^{7} \pm 1.56{\times}10^{7}$
            & $1.62{\times}10^{7} \pm 1.05{\times}10^{7}$ \\
        \bottomrule
    \end{tabular}\\[2pt]
    {\footnotesize
      $^{\dagger}$~Not significant at $\alpha{=}0.05$ (high v0 variance from outlier runs).
    }
\end{table*}


We evaluate four policies across LG4, LG5, LG6, and LG7 scenarios:
(1) LLM (static),
(2) GUIDE (evolved LLM),
(3) Linear Quadratic Regulator (LQR),
and (4) Prograde alignment (see Table~\ref{tab:policy_comparison}).

Performance is measured using the composite interception score:
\begin{equation}
S = d_{\min}^{LB\,2} + \frac{10^{6}}{d_{\min}^{BG} + 0.1},
\end{equation}
where $d_{\min}^{LB}$ and $d_{\min}^{BG}$ denote the minimum lady-bandit and bandit-guard distances. Lower scores are better. Two-sample t-tests ($\alpha=0.05$) compare the baseline LLM and GUIDE.

The evolved LLM substantially improves upon the static LLM baseline across all scenarios (Table~\ref{tab:policy_comparison}), with statistically significant gains in LG4, LG6, and LG7, and a non-significant trend in LG5. Improvements are largest in LG6-LG7, where the guard adopts a defensive position and interception requires conditional maneuvering, while LG4-LG5 primarily reward direct pursuit. Here, simple prograde alignment already provides a strong open-loop interception heuristic, leaving less room for adaptive improvement.
\begin{figure}[ht]
    \centering
    \includegraphics[width=0.85\linewidth]{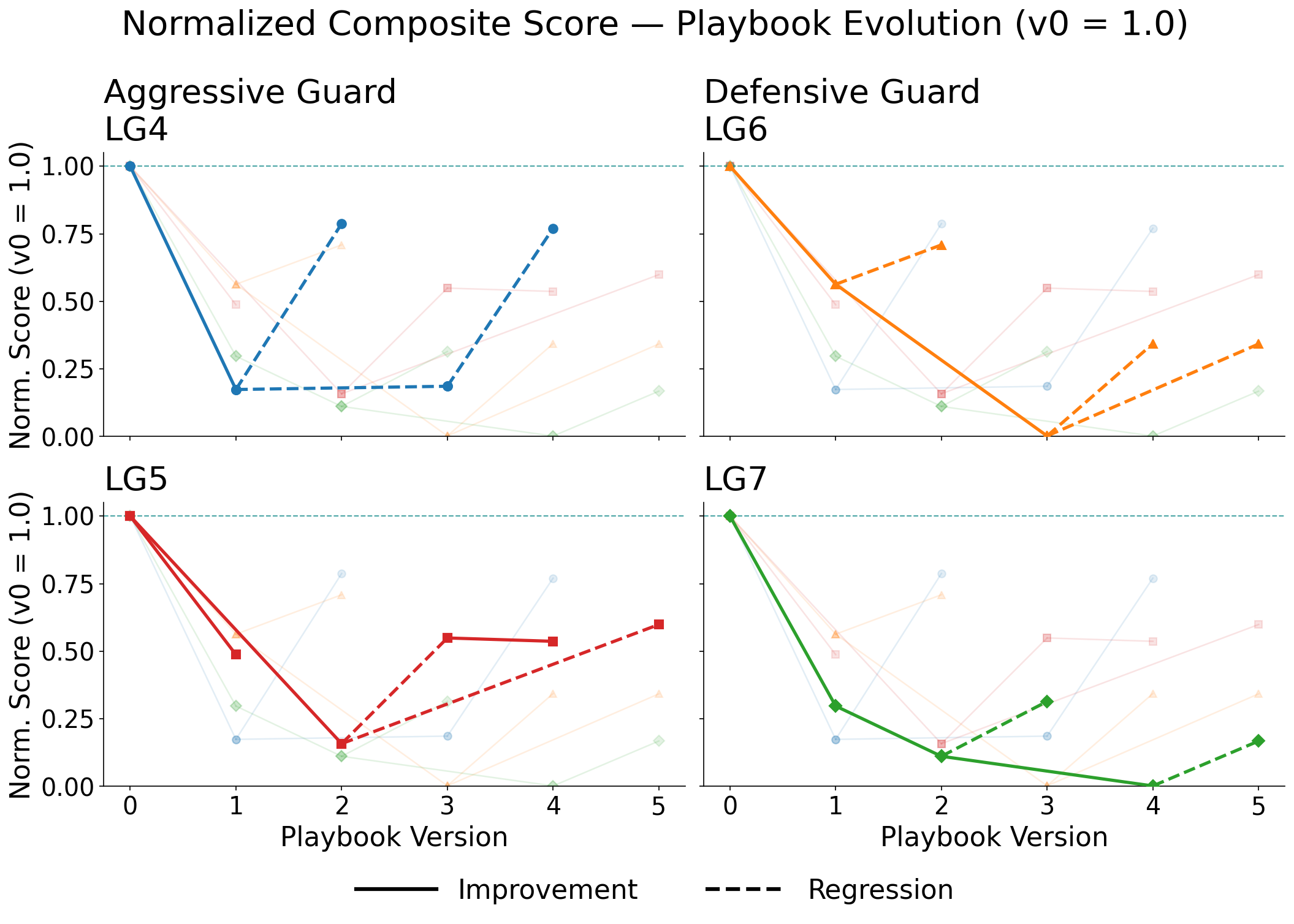}
    \caption{Evolution comparison.}
    \label{fig:evolution_comparison}
\end{figure}

Across all scenarios, the first playbook update consistently identifies the same failure mode: the agent maintains pursuit while the guard closes in to intercept the bandit. Introducing guard-avoidance rules that suspend pursuit and apply lateral thrusts to evade the guard, combined with higher-precision approach behaviors, reduces the composite score by 82\%-99\% within a few evolutions; Table~\ref{tab:distance_breakdown} shows that this arises from both reduced lady-bandit distance and increased guard separation.

\begin{table}[h]
    \centering
    \small
    \setlength{\tabcolsep}{0.8pt}
    \caption{GUIDE distance metrics (m).
             Lady: lady-bandit ($\downarrow$ better).
             Guard: guard-bandit ($\uparrow$ better).}
    \label{tab:distance_breakdown}
    \footnotesize
    \begin{tabular}{lcccccccc}
        \toprule
        & \multicolumn{2}{c}{LG4}
        & \multicolumn{2}{c}{LG5}
        & \multicolumn{2}{c}{LG6}
        & \multicolumn{2}{c}{LG7} \\
        \cmidrule(r){2-3}\cmidrule(r){4-5}\cmidrule(r){6-7}\cmidrule(r){8-9}
        Policy
        & Lady & Guard
        & Lady & Guard
        & Lady & Guard
        & Lady & Guard \\
        \midrule
        LLM v0
            & 237.31 & 13.34
            & 310.95 & \textbf{20.91}
            & 291.83 & 8.41
            & 258.20 & 8.07 \\
        GUIDE (best)
            & \textbf{28.14} & \textbf{17.58}
            & \textbf{11.44} & 14.56
            & \textbf{88.69} & \textbf{30.49}
            & \textbf{52.16} & \textbf{29.43} \\
        \bottomrule
    \end{tabular}
\end{table}

\begin{figure}[ht]
    \centering
    \includegraphics[width=1\linewidth]{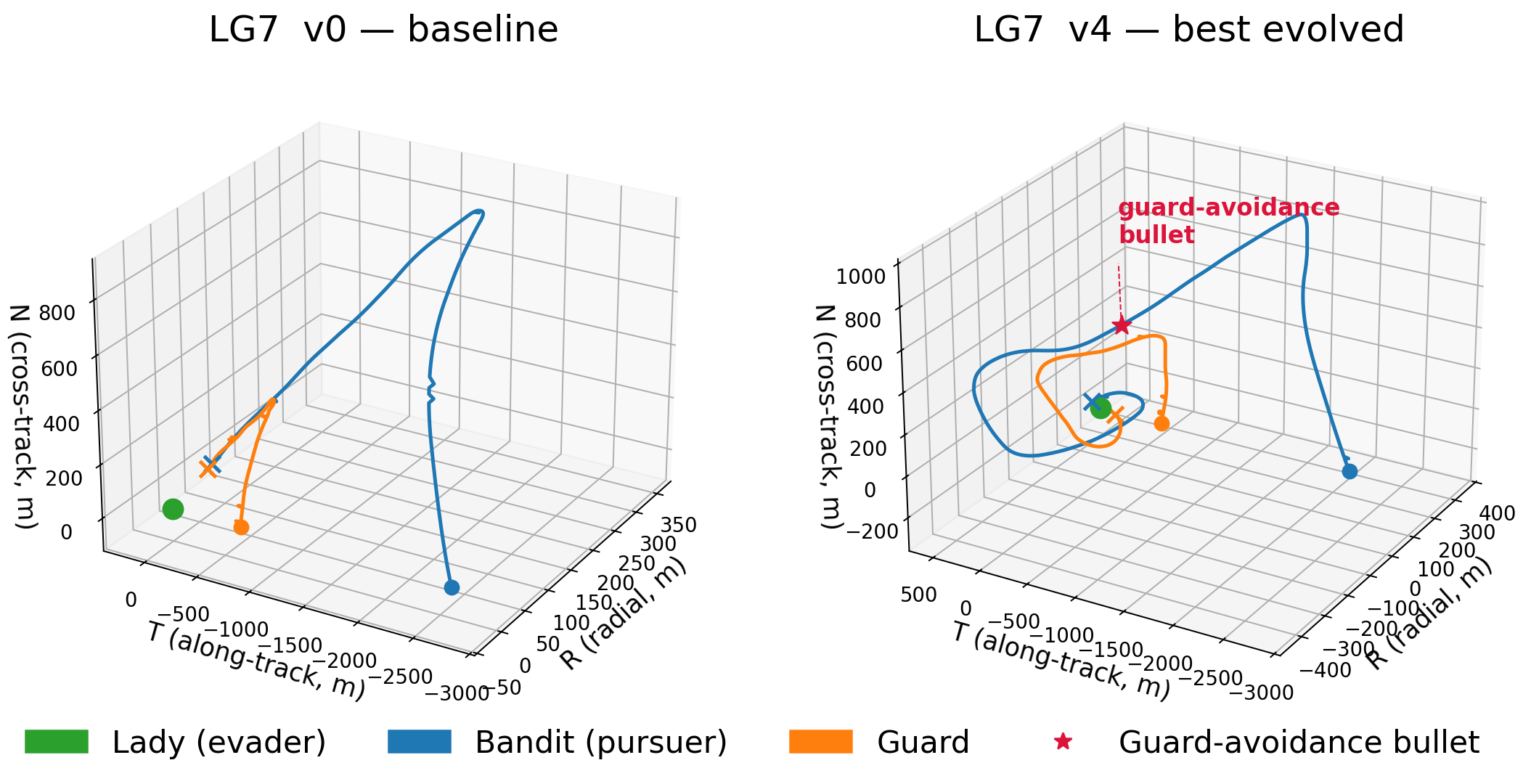}
    \caption{Hill-frame (RTN) trajectories for LG7 (v0 and best).}
    \label{fig:trajectories_evol}
\end{figure}

Improvements arise from structured context updates rather than parameter learning. The offline model encodes corrective strategy into state-conditioned playbook rules executed by the fixed acting model, functioning as a form of in-context distillation that reshapes $\pi(a \mid s, P)$. As shown in Fig.~\ref{fig:trajectories_evol}, activating a single guard-avoidance bullet changes the trajectory, suspending pursuit and inducing lateral escape before re-engagement.

%% file: sec/5_conclusions_and_future_work.tex
\section{Conclusions \& Future Work}

GUIDE demonstrates that LLMs can adapt online to dynamic adversarial scenarios through structured context evolution, without weight updates. Beyond static spacecraft operation, context evolution enables cross-episode policy improvement under unpredictable agent interactions, which is most impactful when baseline policies are far from optimal.

Future work includes ablation studies, tighter low-level control integration, algorithmic candidate evolution alongside language policies, and larger-scale playbook evolution to assess convergence behavior.

%% file: sec/6_acknowledgements.tex
\section{Acknowledgements}
Research was sponsored by the Department of the Air Force Artificial Intelligence Accelerator and was accomplished under Cooperative Agreement Number FA8750-19-2-1000. The views and conclusions contained in this document are those of the authors and should not be interpreted as representing the official policies, either expressed or implied, of the Department of the Air Force or the U.S. Government. The U.S. Government is authorized to reproduce and distribute reprints for
Government purposes notwithstanding any copyright notation herein.
This work has been partially supported by the Spanish Agencia Estatal de Investigacion (AEI) under grant PID2024-161963OB-C22 (ACTIVATION project)

%% file: sec/X_suppl.tex
\clearpage
\setcounter{page}{1}
\maketitlesupplementary
\section{Per-Version Performance Statistics}
\label{supp:perf_by_version}

\subsection{LG4 — Passive Lady, Active Guard}
\begin{table}[H]
  \centering\small
  \caption{LG4 per-version statistics.}
  \label{tab:app_lg4}
  \setlength{\tabcolsep}{3pt}
  \begin{tabular}{crcc}
    \toprule
    Version & Mean score
             & $\bar{d}_{\mathrm{Lady}}$ (m) & $\bar{d}_{\mathrm{Guard}}$ (m) \\
    \midrule
    v0              & $4.15{\times}10^{5}$ & 237.3 & 13.3 \\
    \textbf{v1}     & $\mathbf{7.22{\times}10^{4}}$ & \textbf{28.1} & \textbf{17.6} \\
    v2              & $3.27{\times}10^{5}$ & 192.7 & 15.9 \\
    v3              & $7.72{\times}10^{4}$ & 36.0  & 18.4 \\
    v4              & $3.19{\times}10^{5}$ & 173.9 & 16.7 \\
    \bottomrule
  \end{tabular}
\end{table}

\subsection{LG5 — Passive Lady, Faster Active Guard}
\begin{table}[H]
  \centering\small
  \caption{LG5 per-version statistics.}
  \label{tab:app_lg5}
  \setlength{\tabcolsep}{3pt}
  \begin{tabular}{crcc}
    \toprule
    Version & Mean score
             & $\bar{d}_{\mathrm{Lady}}$ (m) & $\bar{d}_{\mathrm{Guard}}$ (m) \\
    \midrule
    v0              & $5.37{\times}10^{5}$ & 311.0 & 20.9 \\
    v1              & $2.62{\times}10^{5}$ & 159.7 & 21.0 \\
    \textbf{v2}     & $\mathbf{8.49{\times}10^{4}}$ & \textbf{11.4} & \textbf{14.6} \\
    v3              & $2.95{\times}10^{5}$ & 164.0 & 19.8 \\
    v4              & $2.88{\times}10^{5}$ & 164.5 & 38.0 \\
    v5              & $3.21{\times}10^{5}$ & 160.5 & 13.3 \\
    \bottomrule
  \end{tabular}
\end{table}

\subsection{LG6 — Passive Lady, Blocking Guard}
\begin{table}[H]
  \centering\small
  \caption{LG6 per-version statistics.}
  \label{tab:app_lg6}
  \setlength{\tabcolsep}{1.25pt}
  \begin{tabular}{crcc}
    \toprule
    Version & Mean score
             & $\bar{d}_{\mathrm{Lady}}$ (m) & $\bar{d}_{\mathrm{Guard}}$ (m) \\
    \midrule
    v0              & $2.28{\times}10^{7}$ & 291.8 & 8.4  \\
    v1              & $1.29{\times}10^{7}$ & 164.8 & 9.0  \\
    v2              & $1.62{\times}10^{7}$ & 254.7 & 13.5 \\
    \textbf{v3}     & $\mathbf{5.07{\times}10^{4}}$ & \textbf{88.7} & \textbf{30.5} \\
    v4              & $7.85{\times}10^{6}$ & 106.3 & 13.2 \\
    v5              & $7.82{\times}10^{6}$ & 123.1 & 11.5 \\
    \bottomrule
  \end{tabular}
\end{table}

\subsection{LG7 — Passive Lady, Dual Guard}
\begin{table}[H]
  \centering\small
  \caption{LG7 per-version statistics.}
  \label{tab:app_lg7}
  \setlength{\tabcolsep}{1.25pt}
  \begin{tabular}{crrcc}
    \toprule
    Version & $n$ & Mean score
             & $\bar{d}_{\mathrm{Lady}}$ (m) & $\bar{d}_{\mathrm{Guard}}$ (m) \\
    \midrule
    v0              & 20 & $1.95{\times}10^{7}$ & 258.2 & 8.1  \\
    v1              & 10 & $5.79{\times}10^{6}$ & 57.4  & 11.9 \\
    v2              & 15 & $2.17{\times}10^{6}$ & 59.0  & 24.4 \\
    v3              &  5 & $6.13{\times}10^{6}$ & 55.7  & 13.8 \\
    \textbf{v4}     & \textbf{20} & $\mathbf{3.49{\times}10^{4}}$ & \textbf{52.2} & \textbf{29.4} \\
    v5              &  5 & $3.27{\times}10^{6}$ & 142.0 & 38.0 \\
    \bottomrule
  \end{tabular}
\end{table}

\begin{figure*}[t]
  \centering
  \includegraphics[width=\textwidth]{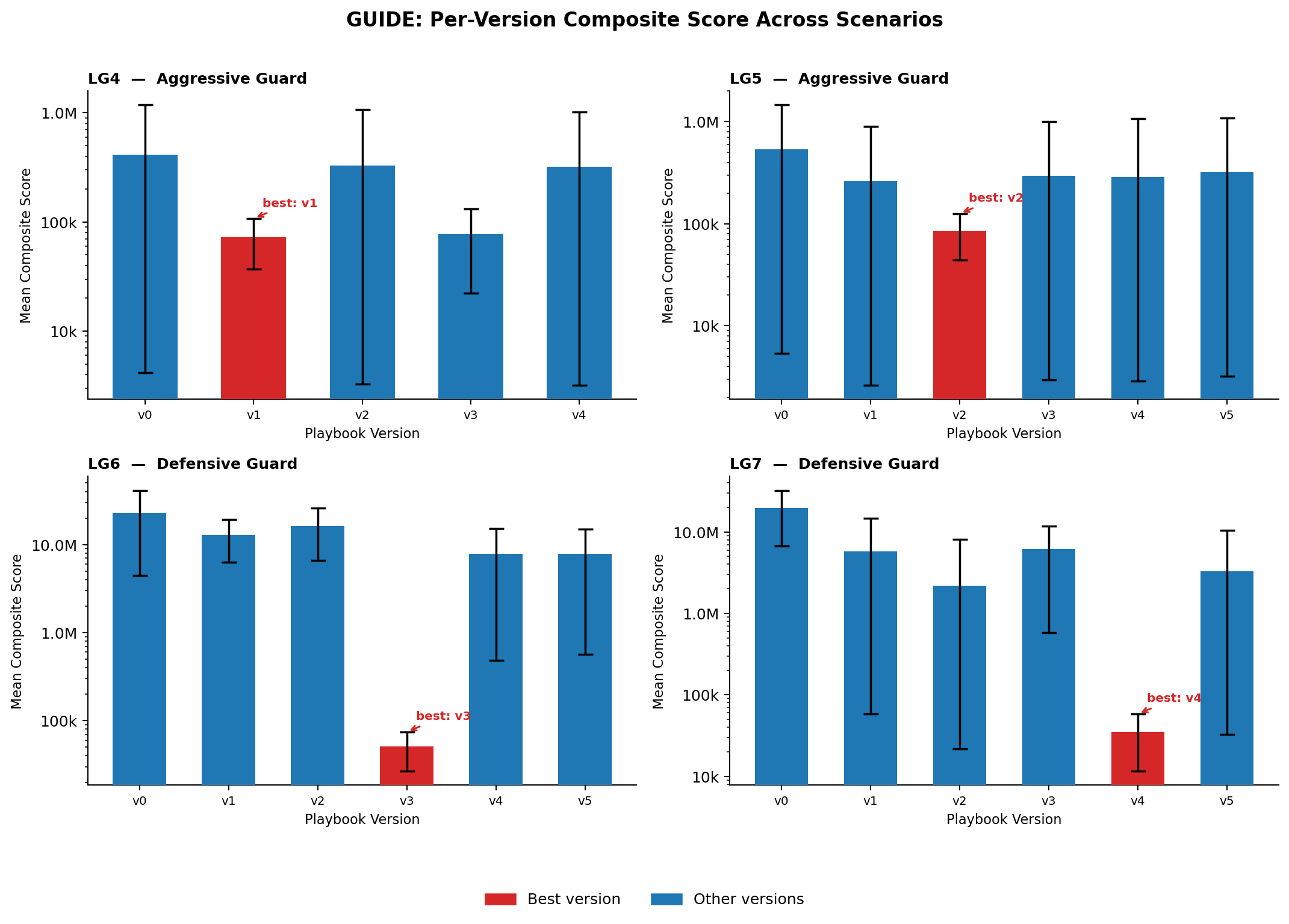}
  \caption{Per-version mean composite score (logarithmic scale, $\pm$1\,std) for all four scenarios. Red bars denote the best-performing version.}
  \label{fig:evolution_curves}
\end{figure*}

\newpage

\section{GUIDE Playbook Structure and Examples}
\label{supp:playbooks}

\begin{figure*}[htbp]
\centering
\fbox{
\begin{minipage}{0.96\textwidth}
\ttfamily\small
id:\quad\quad\quad\quad\quad <unique id, e.g.\ guard-avoidance-00001>\\
section:\quad\quad\quad <guard\_avoidance \textbar{} approach \textbar{} \ldots>\\
type:\quad\quad\quad\quad <constraint \textbar{} rule>\\
text:\quad\quad\quad\quad <NL instruction, 1-3 sentences>\\
conditions:\\
\quad time:\quad\quad\quad\quad\quad\{min: <seconds>\}\quad\quad \textrm{// ignore early orbit phase}\\
\quad guard\_distance: \{max: <metres>\}\quad \textrm{// guard proximity trigger}\\
\quad guard\_approaching: <bool>\quad\quad\quad \textrm{// guard closing flag}\\
\quad target\_distance: \{max: <metres>\}\quad \textrm{// optional: Lady proximity}\\
\quad velocity:\quad\quad\quad\{max: <m/s>\}\quad\quad \textrm{// optional: approach speed}\\
states:\quad\quad\quad\quad <list of (t,\ d\_lady,\ v,\ d\_guard) snapshots>\\
evidence:\quad\quad\quad <causal narrative written by LLM from failure episode>\\
episode\_history: [<episode ids that refined this bullet>]\\
occurrence\_count: <int>\quad\quad \textrm{// times this rule fired}
\end{minipage}
}
\caption{GUIDE playbook bullet schema. The \texttt{conditions} block is a
  \emph{symbolic guard}: the bullet text is injected into the LLM prompt only
  when all conditions evaluate to true on the current observation.}
\label{fig:bullet_schema}
\end{figure*}

\begin{figure*}[htbp]
\centering
\fbox{
\begin{minipage}{0.96\textwidth}
\small
\textbf{id:}~\texttt{guard-avoidance-00001}\quad
\textbf{section:}~\texttt{guard\_avoidance}\quad
\textbf{type:}~\texttt{constraint}\quad
\textbf{occurrence\_count:}~1\\[5pt]

\textbf{text:}~\textit{``When the Guard is closing inside $\sim$220\,m, stop all
forward pursuit and instead apply continuous lateral and/or vertical evasive
thrust (right/left and up/down only) until the Guard distance increases above
this threshold.''}\\[5pt]

\textbf{conditions:}~\texttt{time.min=35\,s},~\texttt{guard\_distance.max=220\,m},~\texttt{guard\_approaching=true}\\[5pt]

\textbf{evidence:}~\textit{Guard closed from $\sim$230\,m at $t\approx$125.4\,s
to 16.9\,m at $t=137.0$\,s while forward throttle (\texttt{fx=1.0}) was
maintained, causing a proximity violation with \texttt{guard\_distance=30.1\,m}.}
\end{minipage}
}
\caption{\textbf{Example 1} — Simple \texttt{guard\_avoidance} constraint, LG6~v2.
  Produced after Episode~2 where the Guard closed from 230\,m to 17\,m in 11.5\,s
  while the Bandit maintained forward throttle.}
\label{fig:bullet_ex1}
\end{figure*}

\begin{figure*}[htbp]
\centering
\fbox{
\begin{minipage}{0.96\textwidth}
\small
\textbf{id:}~\texttt{guard-avoidance-00001}\quad
\textbf{section:}~\texttt{guard\_avoidance}\quad
\textbf{type:}~\texttt{constraint}\quad
\textbf{occurrence\_count:}~3\\[5pt]

\textbf{text:}~\textit{``After the initial phase ($t\geq$35\,s), apply a
two-tiered guard-avoidance regime:}

\textit{\textbf{(1) Caution zone} (\texttt{guard\_distance}~$\lesssim$~230\,m):
Immediately stop using any forward or backward thrust that further reduces Guard
range relative to the Lady pursuit path.
Set longitudinal throttle to neutral (\texttt{fx=0}) unless a brief backward input
is clearly increasing \texttt{guard\_distance}.
Prioritise lateral and vertical evasive thrust (right/left, up/down only) to start
opening separation from the Guard.}

\textit{\textbf{(2) Critical zone} (\texttt{guard\_distance}~$\lesssim$~180\,m,
urgency at $\lesssim$~160\,m):
Immediately freeze all forward and backward throttle (\texttt{ft=0}), overriding
any Lady-approach logic.
Focus exclusively on lateral and vertical manoeuvres until \texttt{guard\_distance}
reopens beyond $\sim$230\,m.}

\textit{Do not resume forward pursuit of the Lady until the Guard is no longer
approaching and \texttt{guard\_distance} has opened beyond $\sim$230-260\,m.''}\\[5pt]

\textbf{conditions:}~\texttt{time.min=35\,s},~\texttt{guard\_distance.max=230\,m},~\texttt{guard\_approaching=true},
~\texttt{target\_distance.max=900\,m},~\texttt{velocity.max=25\,m/s}\\[5pt]

\textbf{evidence:}~\textit{In one case, the Guard closed from $\sim$300\,m to 86.1\,m while strong forward/braking inputs maintained a nose-to-Lady approach, causing a proximity violation. In another, the Guard closed from $\sim$270\,m to 15.5\,m while forward throttle was held nearly continuously, causing a critical proximity violation. Both show that continuing longitudinal motion toward the Lady while the Guard is closing inside a few hundred metres leads to unsafe proximity.}
\end{minipage}
}
\caption{\textbf{Example 2} — Tiered \texttt{guard\_avoidance} constraint, LG7~v4
  (\texttt{episode\_history}: episodes 1, 2, 4).
  The two-zone structure emerged across three failure episodes; each zone threshold
  was calibrated by a different episode and synthesised into a single protocol.}
\label{fig:bullet_ex2}
\end{figure*}

\begin{figure*}[htbp]
\centering
\fbox{
\begin{minipage}{0.96\textwidth}
\small
\textbf{id:}~\texttt{approach-00002}\quad
\textbf{section:}~\texttt{approach}\quad
\textbf{type:}~\texttt{constraint}\quad
\textbf{occurrence\_count:}~2\\[5pt]

\textbf{text:}~\textit{``When within $\sim$120\,m of the Lady and still approaching
faster than $\sim$18\,m/s, switch to backward throttle or zero forward throttle
until relative speed is reduced below 18\,m/s to avoid overshoot.''}\\[5pt]

\textbf{conditions:}~\texttt{target\_distance.max=120\,m},~\texttt{velocity.min=18\,m/s},~\texttt{approaching=true},~\texttt{time.min=35\,s}\\[5pt]

\textbf{evidence:}~\textit{Bandit was 76\,m from Lady at 23\,m/s with no braking applied, resulting in overshoot.}
\end{minipage}
}
\caption{\textbf{Example 3} — \texttt{approach} braking constraint, LG4~v2.
  Addresses terminal-phase kinematics when guard-avoidance permits Lady approach
  but closure speed is too high.}
\label{fig:bullet_ex3}
\end{figure*}